\documentclass[sigconf, nonacm]{acmart}

\usepackage{subcaption}
\usepackage{tabularx}
\usepackage{multirow}

\newcommand{\smalltt}[1]{{\texttt{\small #1}}}

\newcolumntype{L}[1]{>{\raggedright\let\newline\\\arraybackslash\hspace{0pt}}m{#1}}
\newcolumntype{C}[1]{>{\centering\let\newline\\\arraybackslash\hspace{0pt}}m{#1}}
\newcolumntype{R}[1]{>{\raggedleft\let\newline\\\arraybackslash\hspace{0pt}}m{#1}}

\definecolor{green}{RGB}{28,171,70}
\definecolor{yellow}{RGB}{218,187,56}

\newcommand{\mlarge}{\texttt{page\_leap()}}
\newcommand{\m}{\smalltt{page\_leap()}}

\newif\ifshowreviewtext
\showreviewtexttrue   

\showreviewtextfalse

\newcommand{\refReviewW}[2]{\noindent\textbf{\hyperref[R#1O#2]{R#1O#2}}}
\newcommand{\refReviewD}[2]{\noindent\textbf{\hyperref[R#1D#2]{R#1D#2}}}

\begin{document}

\title{Taking the Leap: Efficient and Reliable Fine-Grained NUMA Migration in User-space}

\author{Felix Schuhknecht}
\affiliation{%
  \institution{Johannes Gutenberg University}
  \city{Mainz, Germany}
  \country{}}
\email{schuhknecht@uni-mainz.de}

\author{Nick Rassau}
\affiliation{%
  \institution{Johannes Gutenberg University}
  \city{Mainz, Germany}
  \country{}}
\email{rassau@uni-mainz.de}

\renewcommand{\shortauthors}{Felix Schuhknecht and Nick Rassau}

\begin{abstract}
\vspace*{-0.2cm}
Modern multi-socket architectures offer a single virtual address space, but physically divide main-memory across multiple regions, where each region is attached to a CPU and its cores. While this simplifies the usage, developers must be aware of non-uniform memory access (NUMA), where an access by a thread running on a core-local NUMA region is significantly cheaper than an access from a core-remote region. 
Obviously, if query answering is parallelized across the cores of multiple regions, then the portion of the database on which the query is operating should be distributed across the same regions to ensure local accesses. As the present data placement might not fit this, migrating pages from one NUMA region to another can be performed to improve the situation. 
To do so, different options exist: One option is to rely on  automatic NUMA balancing integrated in Linux, which is steered by the observed access patterns and frequency. Another option is to actively trigger migration via the system call \smalltt{move\_pages()}. Unfortunately, both variants have significant downsides in terms of their feature set and performance. 
As an alternative, we propose a new user-space migration method called \m{} that can perform page migration asynchronously at a high performance by exploiting features of the virtual memory subsystem. The method is (a)~actively triggered by the user, (b)~ensures that all pages are eventually migrated, (c)~handles concurrent writes correctly, (d)~supports pooled memory, (e)~adaptively adjusts its migration granularity based on the workload, and (f)~supports both small pages and huge pages. 

\end{abstract}

\maketitle

\vspace*{-0.4cm}
\section{Introduction}
\vspace*{-0.1cm}

Multi-socket server architectures offer a single virtual address space, but physically divide main-memory across multiple regions, where each region is attached to a CPU and its cores. While this single virtual address space simplifies development, programmers must still be cautious: If a thread 
accesses physical memory on a NUMA region that is different from the one of the core on which it currently runs, then this \textit{remote access} is significantly more expensive than a \textit{local access}. Consequently, these so-called \textit{non-uniform memory accesses (NUMA)}~\cite{lit:numa} must be taken into account when aiming for high performance. 
In DBMSs, if query answering is parallelized across the cores of multiple regions, then the portion of the database on which the query is operating should be distributed across the same regions to ensure local accesses. Unfortunately, the current data placement might not fit this parallelization. To improve upon the situation, individual physical memory pages can be \textit{migrated} from one NUMA region to another at runtime to bring them closer to the worker thread and thus speed up future accesses. 

To perform migration, there are two different options: 
(1)~To rely on the \textit{implicit page migration} mechanism of the OS, which is carried out automatically and transparently by the kernel. The core concept is that the OS monitors the number of accesses and access patterns on allocated memory regions at a certain frequency. If it observes a sufficient amount of remote accesses, it might decide to migrate pages to a different region. Of course, migrating pages is costly, but it might pay off in the future, so the heuristic aims at balancing this tradeoff. Linux calls the concept of implicit page migration \textit{auto NUMA balancing}~\cite{lit:auto_numa_balancing}, which on many distributions is activated by default (\smalltt{/proc/sys/kernel/numa\_balancing} set to \smalltt{1}) and works for anonymous memory allocations backed by small pages and transparent huge pages, e.g., as created by \smalltt{mmap()}. 
While automatic NUMA balancing works transparently, it has two downsides: 
(a)~The user does not have any control over when page migration happens and whether it happens at all. As our evaluation will show, frequent concurrent modifications, as they are common in DBMSs facing transactional workloads, make auto NUMA balancing often ``wait'' for times of little load to do the migration, which might actually never come. This leads to continuous stream of unnecessary remote accesses.
Also, (b)~the migration is performed into freshly allocated memory and does not allow migration into already allocated, pooled memory. This causes costly page faults, rendering the migration unnecessarily expensive if the application anyways pools memory, like DBMSs often do. 
These problems lead us to the second type of page migration mechanism, namely (2)~\textit{explicit page migration} methods. In this case, the migration of pages is actively triggered by the programmer by calling a corresponding function and by specifying the pages to migrate. In Linux, this migration function is the ancient \smalltt{move\_pages()} system call, which is wrapped by the \smalltt{numa\_move\_pages()} method~\cite{lit:move_pages} from \smalltt{libnuma}. It allows to migrate a specified subset of the currently allocated pages from user-space. 
While this system call provides control over the migration, it unfortunately also comes with problems: 
(a)~There is no option to adjust the migration granularity; it always happens at page size. If a large number of pages must be migrated, this can heavily penalize the calling thread.
(b)~There is still no guarantee that the page migration of all pages is performed. If a page is currently busy (i.e. the subsystem is holding a reference to the page) or the page became dirty during migration, it could simply not migrate it.
(c)~Just like auto NUMA balancing, it does not allow migration into pooled memory and hence suffers from page faults. 

\vspace*{-0.2cm}
\subsection{User-space Remapping: \mlarge{}}
\vspace*{-0.1cm}

Ideally, we would like to have a migration method that is controllable from user-space, migrates all requested pages with certainty, and supports migration into pooled memory. Further, the method should be thread-safe and tolerate concurrent modifications. 
In the following, we propose exactly such a method, which we call \textit{\m{}}. The core concept is based on the technique of memory rewiring~\cite{lit:rewiring} by which the mapping from virtual to physical memory can be manipulated from user-space at runtime. To perform a page migration, we carry out two steps: one on the physical level and one on the virtual level. On the physical level, the page content is non-atomically copied from the source NUMA region to the destination NUMA region, potentially into already-pooled memory. On the virtual level, the virtual page is then atomically remapped to the physical page of the destination region, rendering it ``active''. To detect and handle concurrent writes that would invalidate an ongoing page migration, we use a segmentation fault handler that invalidates the migration of the page and triggers a new migration attempt until eventually all pages have been migrated. Our approach also supports different migration granularities that are adaptively adjusted based on the observed workload.
Table~\ref{tab:feature_comparison} summarizes the features of \m{} in comparison with auto NUMA balancing and \smalltt{move\_pages()}.  

\begin{table}[h!]
\vspace*{-0.3cm}
\begin{center}
\footnotesize
\begin{tabular}{C{2.7cm}|c|c|c|c} 
 \toprule
 \textbf{Method} & \textbf{Pooling}  & \textbf{Controlled} & \textbf{Reliable} & \textbf{Adaptive}\\ \midrule
 Auto NUMA balancing & \textcolor{red}{$\times$} & \textcolor{red}{$\times$} & \textcolor{red}{$\times$} & \textcolor{red}{$\times$} \\\hline
 \smalltt{move\_pages()} & \textcolor{red}{$\times$} & \textcolor{green}{$\checkmark$} & \textcolor{red}{$\times$} & \textcolor{red}{$\times$}\\\hline
\m{} & \textcolor{green}{$\checkmark$} & \textcolor{green}{$\checkmark$} & \textcolor{green}{$\checkmark$} & \textcolor{green}{$\checkmark$} \\
 \bottomrule
\end{tabular}
\end{center}
\caption{Feature comparison of auto NUMA balancing, \smalltt{move\_pages()}, and our proposed method \m{}.}
\label{tab:feature_comparison}
\vspace*{-0.6cm}
\end{table}


\vspace*{-0.2cm}
\subsection{Experimental Environment and Setup}
\label{ssec:setup}
\vspace*{-0.1cm}

For the upcoming experimental evaluation, we use a two-socket server equipped with two Intel Xeon Gold 6326 CPUs running Alma Linux with a vanilla kernel 5.14. Each NUMA region has $128$GB of RAM, resulting in $256$GB in total. Auto NUMA balancing is activated by setting \smalltt{/proc/sys/kernel/numa\_balancing} to \smalltt{1}) The code is compiled with g++11.4 on optimization level~O3. All reported runtimes are the average of three runs. 
For all upcoming experiments, we pin the main thread of the process, which performs all measured accesses to the memory, to the first core of NUMA region~1 using the \smalltt{taskset -c 1} command~\cite{lit:taskset}. This means that NUMA region~0 will be the remote region, while NUMA region~1 will be the local region in the following. To use huge pages for~\m{}, a \smalltt{hugetlbfs} filesystem must be mounted and a sufficient number of huge pages on each NUMA region must be pre-allocated. The same holds for \smalltt{move\_pages()}, which we extended accordingly. Auto NUMA balancing relies on transparent huge pages, as it does not support migrating \smalltt{hugetlbfs} huge pages.

\vspace*{-0.2cm}
\section{Local Accesses vs Remote Accesses}
\label{sec:local_vs_remote_accesses}
\vspace*{-0.1cm}

We start by analyzing how much more expensive remote accesses actually are over local accesses on the present machine. To do so, we allocate a memory area of $4$GB on NUMA region~0 (remote) respectively NUMA region~1 (local) using the \smalltt{mbind()} method which permanently binds the memory to the corresponding region. After initializing the memory with random integers, we perform a set of accesses on the region from our main thread running on NUMA region~1. We test four different access patterns (sequential read/writes and random read/writes) to see whether the pattern has an impact on the cost. For sequential reads/writes, we simply access each byte from the start of the area to the end, resulting in around $4.2$~billion sequential accesses in total. For random reads/writes, we access $10$M randomly selected bytes.
\begin{figure}[h!]
  \centering
  \begin{subfigure}[b]{.49\linewidth}
    \includegraphics[width=\linewidth, trim={0 0 0 0}, clip]{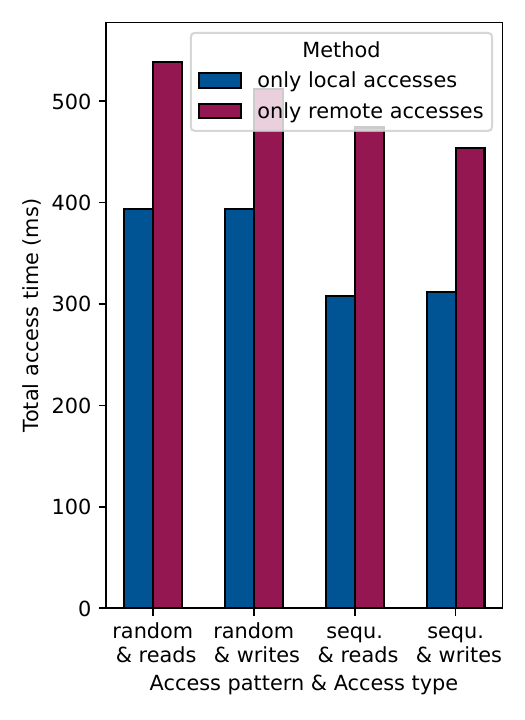}
    \caption{Small pages.}
    \label{fig:local_vs_remote:random}
  \end{subfigure}
  \begin{subfigure}[b]{.49\linewidth}
       \includegraphics[width=\linewidth, trim={0 0 0 0}, clip]{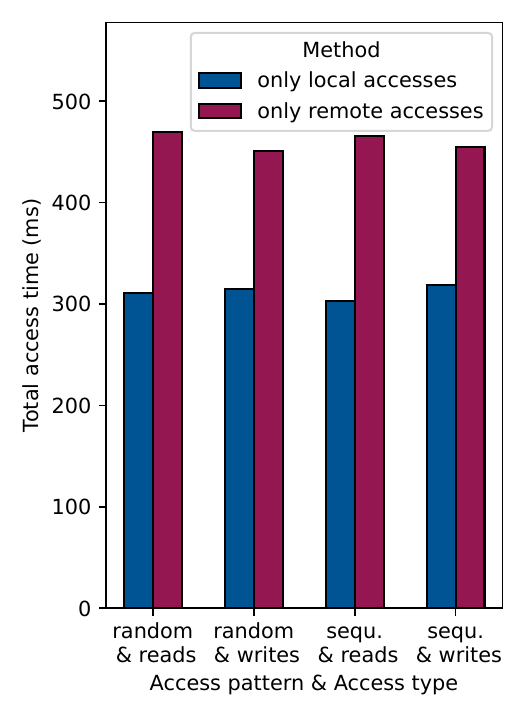}
    \caption{Huge pages.}
    \label{fig:local_vs_remote:sequential}
  \end{subfigure}
    \vspace*{-0.2cm}
  \caption{Local accesses vs remote accesses under different access patterns for small pages and huge pages.}
    \label{fig:local_vs_remote}
    \vspace*{-0.3cm}
\end{figure}
Figure~\ref{fig:local_vs_remote} shows the results for both small pages and huge pages. 
We can observe that for all access patterns, remote accesses are indeed significantly more expensive than local accesses, justifying the need for local accesses and hence efficient NUMA migration. Interestingly, the difference between remote and local accesses is for sequential accesses as pronounces as for random accesses, indicating that prefetching under a sequential access pattern cannot hide the cost of remote accesses. 

\vspace*{-0.2cm}
\section{Page migration using \texttt{move\_pages()}}
\label{sec:move_pages}
\vspace*{-0.1cm}

We next evaluate \smalltt{move\_pages()} for explicit page migration. To asses the overhead of \smalltt{move\_pages()}, we first compare it against simply copying the content of all pages from one NUMA region to the other via traditional \smalltt{memcpy()}. As every migration method has at least to physically copy the content of all pages, this baseline is the optimum any migration method can reach. Here, we differentiate between copying into fresh memory, i.e., the copying is the first access to the memory triggering page-faults, and copying into pooled and hence already page-faulted memory. 
Note that \smalltt{memcpy()} is no proper replacement for \smalltt{move\_pages()} or any other migration method due to two limitations: 
(1)~It results in a new virtual location of the data. Hence, all components using the memory must update their references. In contrast, using a migration method, the migration is completely transparent.
(2)~Concurrent writes by another thread to the ``old'' page during the \smalltt{memcpy()} would violate correctness.

\begin{figure}[h!]
    \centering
    \fcolorbox{blue}{white}{
    \includegraphics[width=.95\columnwidth, trim={0 0.4cm 0.2cm 0}, clip]{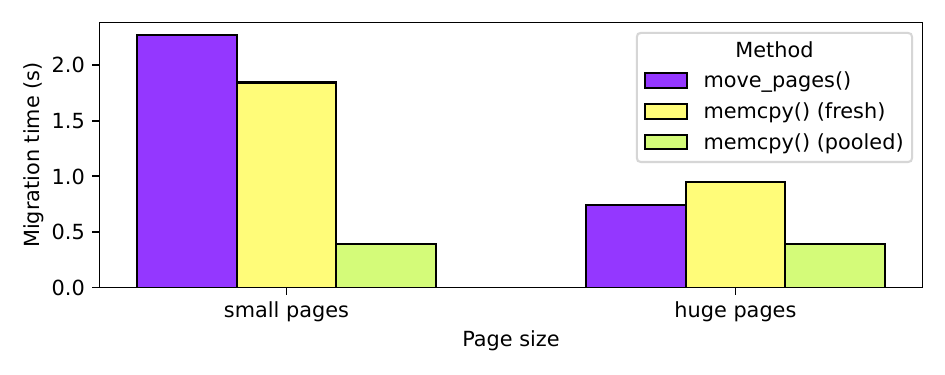}
    }
    \caption{Migration time of \smalltt{move\_pages()} vs \smalltt{memcpy()} from one NUMA region to the other for small and huge pages.}
    \label{fig:move_pages_vs_memcpy}
    \vspace*{-0.3cm}
\end{figure}

Figure~\ref{fig:move_pages_vs_memcpy} shows the results for both small pages and huge pages. We can see that for small pages, the overhead of \smalltt{move\_pages()} over \smalltt{memcpy()} into fresh memory is only $18\%$ while the overhead over \smalltt{memcpy()} into page-faulted memory is a dramatic $82\%$. For huge pages, to our surprise, \smalltt{memcpy()} into fresh memory is actually slightly slower than \smalltt{move\_pages()}, which we attribute to the fact that \smalltt{move\_pages()} runs in a separate thread that is pinned to the destination region. When copying in pooled memory, we also see an overhead of $46$\% for \smalltt{move\_pages()}.
Overall, this shows that for both small pages and huge pages, there is clearly something to gain over \smalltt{move\_pages()} when using pooled memory.

\vspace*{-0.2cm}
\section{\mlarge{} Architecture}
\label{sec:sync_rewiring}
\vspace*{-0.1cm}

The core concept of page migration in \m{} exploits the separation of virtual memory from physical memory in user-space. If the virtual page \smalltt{vpage0} currently maps to a physical page \smalltt{ppage0} that is allocated on NUMA region~0, we essentially perform two simple steps to migrate: First, we allocate a physical page \smalltt{ppage1} on NUMA region~1 from the pool of that region and copy the content from the source physical page to this target physical page using \smalltt{memcpy()}. 
Second, after copying, we remap the virtual page \smalltt{vpage0} to the physical page \smalltt{ppage1} automatically using \smalltt{mmap()}, such that subsequent accesses will hit \smalltt{ppage1} on NUMA region~1. 


To be able to perform such a remapping in the first place, we utilize a technique called memory rewiring~\cite{lit:rewiring, lit:storage_views1, lit:storage_views2}, which introduces a handle to physical memory in user-space in the form of so-called main-memory files. Using \smalltt{mmap()}, virtual pages can be (re)mapped to offsets in the main-memory file, which hence allows to freely adjust the mapping between virtual and physical memory at runtime.  
A nice side-effect of this technique is that the remapping can happen at the granularity of multiple neighboring pages, reducing the number of performed \smalltt{mmap()} calls for areas that are a multiple of the page size.


\vspace*{-0.2cm}
\subsection{Correctly Handling Concurrent Writes}
\vspace*{-0.1cm}

So far, if a write is performed to a portion of the source physical page that has already been copied then after remapping to the target physical page, the write would be lost.   
To handle this situation, we must (a)~detect any write to a area that is currently under copying. If detected, (b)~we must invalidate the copied area and retry the migration. 
\begin{figure}[h!]
  \centering
  \begin{subfigure}[b]{\linewidth}
    \centering
    \includegraphics[page=2, width=.7\linewidth, trim={0 13cm 19cm 0}, clip]{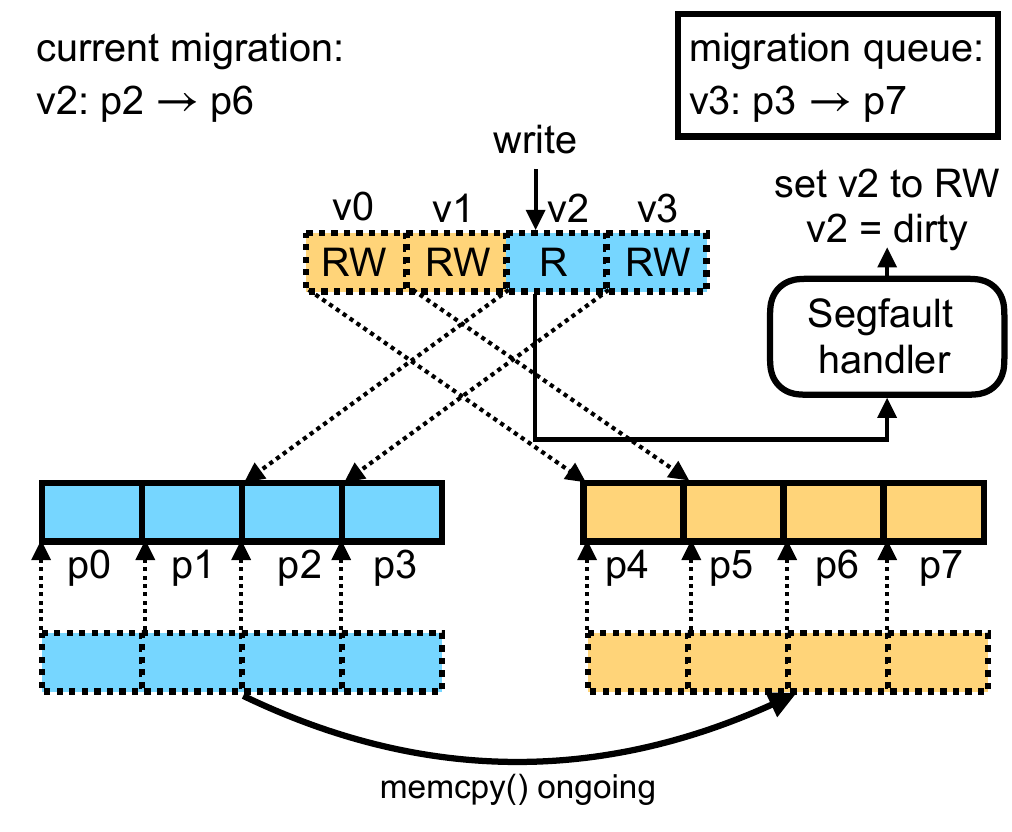}
    \caption{A write to area \smalltt{v2} is happening that is currently being copied. This triggers a segmentation fault, which flags the area as dirty.}
    \label{fig:sync_rewiring_concurrent_writes:first_step}
  \end{subfigure}
  \begin{subfigure}[b]{\linewidth}
    \centering
   \includegraphics[page=3, width=.7\linewidth, trim={0 15cm 19cm 0}, clip]{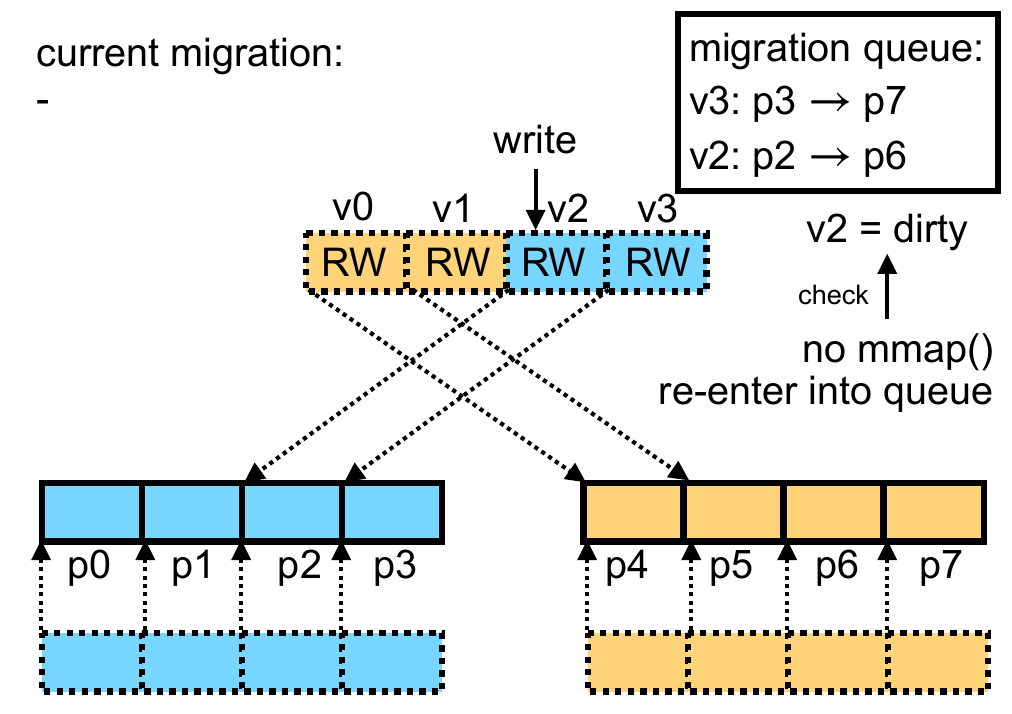}
    \caption{Before the remapping is performed, it is checked whether the area became dirty in the meantime. This is the case, so no remapping is performed, and the area is re-queued for a later migration.}
    \label{fig:sync_rewiring_concurrent_writes:second_step}
  \end{subfigure}
 \vspace*{-0.5cm}
  \caption{Correctly handling concurrent writes for four pages, where third page is currently under migration.}
    \label{fig:sync_rewiring_concurrent_writes}
    \vspace*{-0.4cm}
\end{figure}
To achieve (a)~without forcing the user to actively delegate all writes through a special interface, we use signal handling to do the job. Precisely, we set the protection of the area to migrate to read-only using the \smalltt{mprotect()} system call, as shown in Figure~\ref{fig:sync_rewiring_concurrent_writes:first_step} for the page \smalltt{v2}. By this, a write attempt to the area triggers a segmentation fault, which we can then catch with a custom segmentation fault handler~\cite{lit:anyolap, lit:no_time_to_halt}. This handler is a pre-registered, but otherwise normal user-space function that is called by the OS when the segmentation occurs, and which receives the virtual memory address of the attempted write. In this handler, we then mark the area as \textit{dirty} in a helper structure, reset its protection to writable, and return to retry the write attempt, which now succeeds. 
To take care of (b), when the copying of a area has been performed and is about to become real via remapping, we first check whether it became dirty or not\footnote{In our protocol, it is possible that an area still becomes dirty although its copying has fully happened, namely if a write occurs before the corresponding remapping happens. This rare situation triggers an unnecessary retry of the area, but preserves the write.}, as shown in Figure~\ref{fig:sync_rewiring_concurrent_writes:second_step}. Only if it did not become dirty, we perform the remapping. Otherwise, we insert the area into a queue to retry the migration, where we can adaptively split the area to increase its chance of migration-success as explained in the next section. The process terminates as soon as all pages have been migrated or a specified timeout is reached. Note that this is an advantage over \smalltt{move\_pages()}, which does not guarantee that all pages are migrated.


\vspace*{-0.2cm}
\subsection{Adaptive Area Size}
\vspace*{-0.1cm}

As \m{} can migrate at arbitrary area sizes that are a multiple of the page size, a question is how to chose the area size: A size too small might cause unnecessary overhead due to repetitive required system calls, whereas a size too large increases the likeliness of areas becoming dirty, causing retries. To handle the problem without requiring knowledge about the workload and its pressure in terms of modifications, \m{} implements an adaptive solution: The user must only select an initial area size to start from --- in the experimental evaluation, we will identify such a suitable initial size. Then, during migration, if an area becomes dirty and must be retried, \m{} splits the area into multiple smaller areas based on a reduction factor. By this, \m{} automatically adjusts its area size and hence migration granularity to the pressure of the workload. Additionally, skewed modifications on subsets of the dataset will reduce the area size only for the areas under pressure.



\section{Migration Without Concurrent Writes}
\label{ssec:exp_synchronous}
\vspace*{-0.1cm}

\begin{figure}[h!]
  \centering
  %
  %
  \begin{subfigure}[b]{.48\linewidth}
    \includegraphics[width=\linewidth, trim={0 0 0 0}, clip]{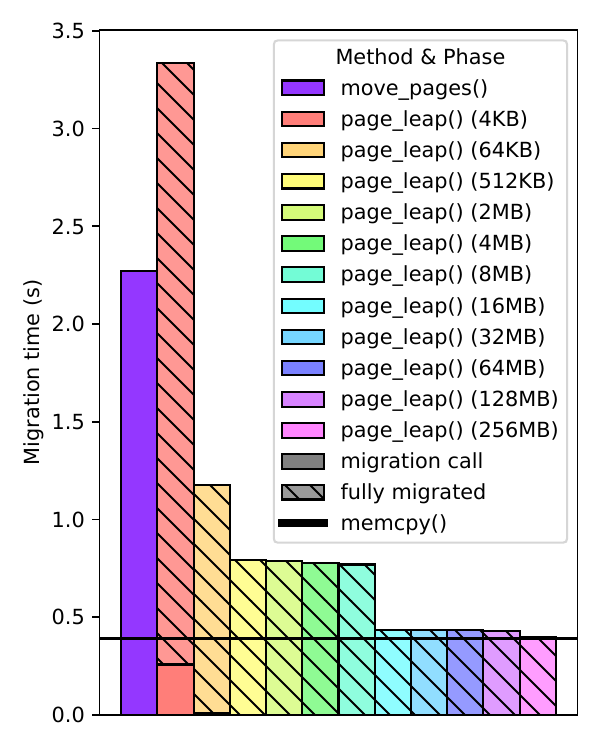}
    \caption{Small pages (pooled).}
    \label{fig:move_pages_vs_sync_rewiring:pooled:random}
  \end{subfigure}
  \begin{subfigure}[b]{.48\linewidth}
    \includegraphics[width=\linewidth, trim={0 0 0 0}, clip]{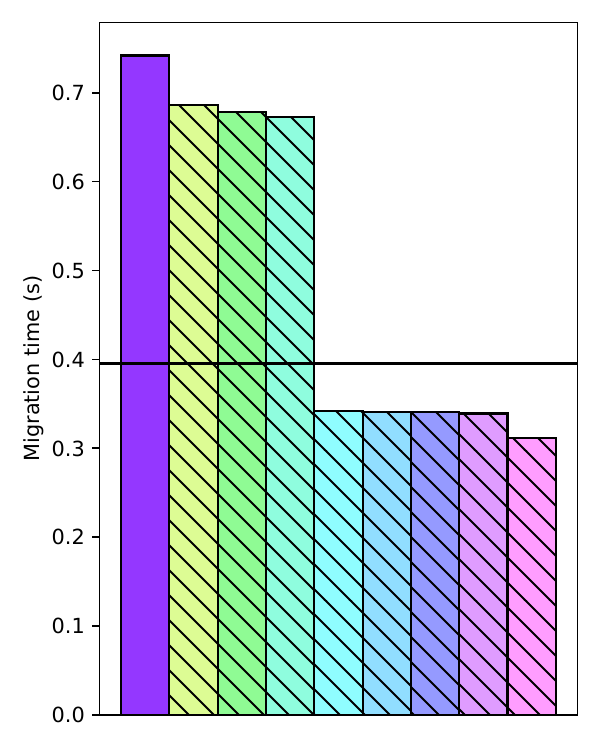}
    \caption{Huge pages (pooled).}
    \label{fig:move_pages_vs_sync_rewiring:pooled:sequential}
  \end{subfigure}

\vspace*{-0.2cm}
  \caption{\smalltt{move\_pages()} vs \m{} without concurrent accesses for different granularities and page sizes. \smalltt{memcpy()} resembles the theoretical optimum.}
    \label{fig:move_pages_vs_sync_rewiring}
    \vspace*{-0.3cm}
\end{figure}

In Figure~\ref{fig:move_pages_vs_sync_rewiring}, we start by evaluating the migration time of \m{} against \smalltt{move\_pages()} for small pages and huge pages without any concurrent accesses happening, resembling the best case for both methods. For both methods, we report the time until the migration is fully completed. For \m{}, we test eleven (small pages) and eight (huge pages) different initial area sizes. As no concurrent accesses are happening in this experiment, no retries are happening and hence no adaptive adjustment of area sizes takes places here. Auto NUMA balancing is not included in this comparison, as there are no accesses that could trigger a migration -- we will include and discuss it in Section~\ref{ssec:exp_asynchronous}. As a baseline, we show the migration time of raw \smalltt{memcpy()} as the theoretical optimum.

\noindent We can see that for both pages sizes, \m{} is able to outperform \smalltt{move\_pages()} in nearly all configurations. Only for the 4KB area size, the system call overhead becomes too dominant. For area sizes of 64KB and 16MB for small pages and huge pages respectively, \m{} outperforms \smalltt{move\_pages()} already by a factor of two or more, demonstrating the superiority of \m{} for migration without concurrent accesses. 
We can also observe plateaus within each plot, showing improvement jumps at $512$KB, and $16$MB. This indicates that these area sizes are very good candidates for the initial area sizes to start from under adaptive splitting. Also, we can observe that \m{} reaches the optimum of \smalltt{memcpy()} for both page sizes at an area size of $16$MB already, showing that \m{} becomes dominated by the copying phase.

\begin{figure*}[h!]
  \centering
  \raisebox{1.1cm}{\rotatebox{90}{\textbf{Migration time}}}
  \begin{subfigure}[b]{.23\linewidth}
  \centering \textbf{10K uniform writes/s}
    \includegraphics[width=\linewidth, trim={0 0.45cm 0 0}, clip]{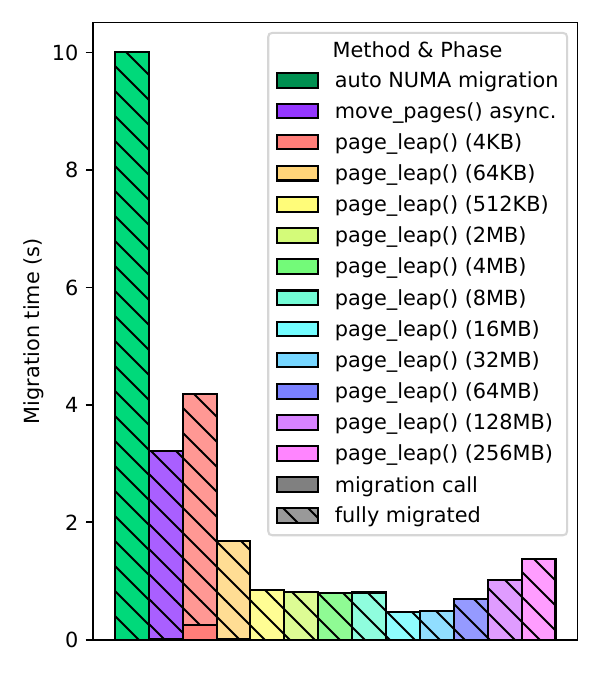}
    \vspace*{-0.3cm}
    \label{fig:async:w1:migration}
  \end{subfigure}
  \begin{subfigure}[b]{.23\linewidth}
  \centering \textbf{100K uniform writes/s}
    \includegraphics[width=\linewidth, trim={0 0.45cm 0 0}, clip]{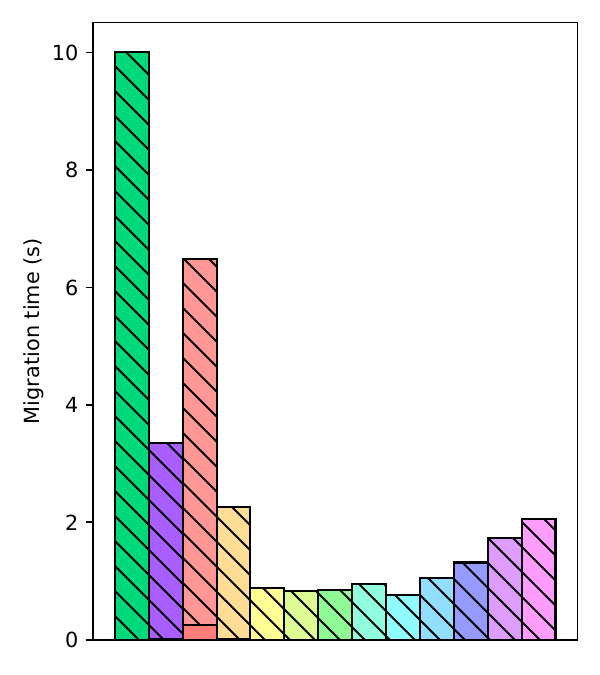}
    \vspace*{-0.3cm}
    \label{fig:async:w2:migration}
  \end{subfigure}
   \begin{subfigure}[b]{.23\linewidth}
     \centering \textbf{10M uniform writes/s}
    \includegraphics[width=\linewidth, trim={0 0.45cm 0 0}, clip]{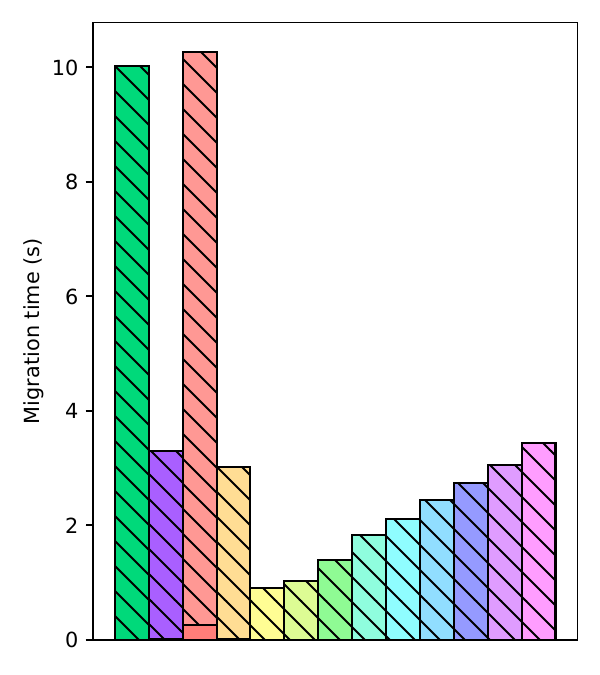}
    \vspace*{-0.3cm}
    \label{fig:async:w3:migration}
  \end{subfigure}
   \begin{subfigure}[b]{.23\linewidth}
    \centering \textbf{100K writes/s, 75\% in 128MB}
    \includegraphics[width=\linewidth, trim={0 0.45cm 0 0}, clip]{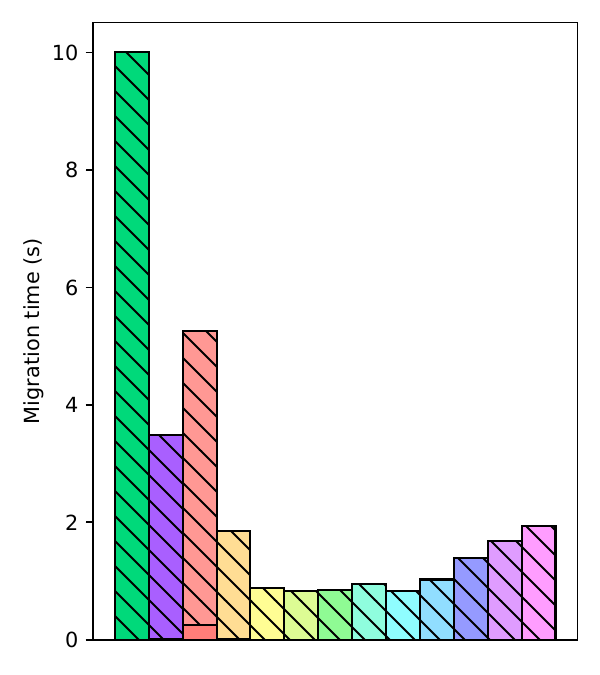}
    \vspace*{-0.3cm}
    \label{fig:async:w4:migration}
  \end{subfigure}

\raisebox{0.1cm}{\rotatebox{90}{\textbf{Throughput}}}
  \begin{subfigure}[b]{.23\linewidth}
    \includegraphics[width=\linewidth, trim={0 0.45cm 0 0}, clip]{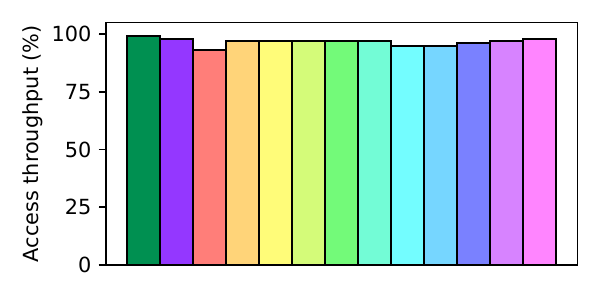}
    \vspace*{-0.3cm}
    \label{fig:async:w1:access_throughput}
  \end{subfigure}
  \begin{subfigure}[b]{.23\linewidth}
    \includegraphics[width=\linewidth, trim={0 0.45cm 0 0}, clip]{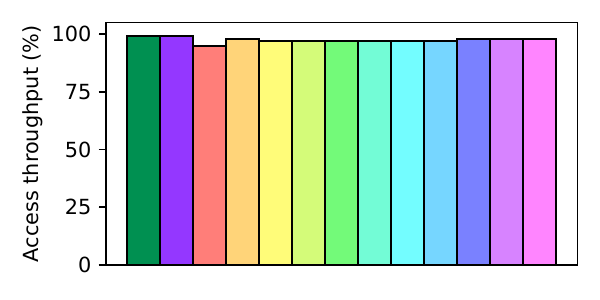}
    \vspace*{-0.3cm}
    \label{fig:async:w2:access_throughput}
  \end{subfigure}
   \begin{subfigure}[b]{.23\linewidth}
    \includegraphics[width=\linewidth, trim={0 0.45cm 0 0}, clip]{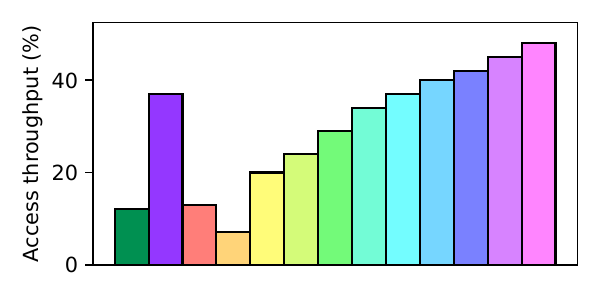}
    \vspace*{-0.3cm}
    \label{fig:async:w3:access_throughput}
  \end{subfigure}
   \begin{subfigure}[b]{.23\linewidth}
    \includegraphics[width=\linewidth, trim={0 0.45cm 0 0}, clip]{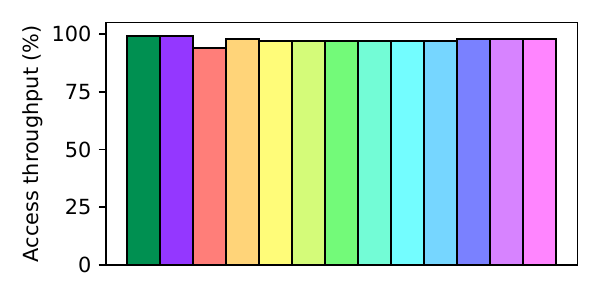}
    \vspace*{-0.3cm}
    \label{fig:async:w4:access_throughput}
  \end{subfigure}  

\raisebox{0.1cm}{\rotatebox{90}{\textbf{Page status}}}
  \begin{subfigure}[b]{.23\linewidth}
    \includegraphics[page=1, width=\linewidth, trim={0 22.3cm 26.1cm 0}, clip]{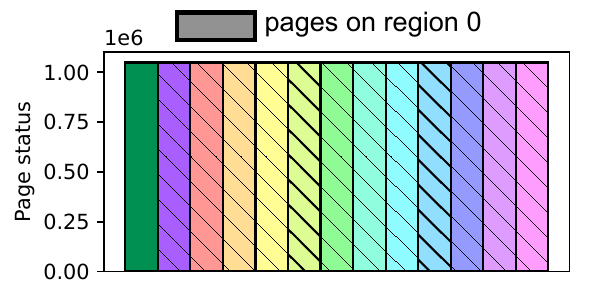}
    \vspace*{-0.3cm}
    \label{fig:async:w1:page_status}
  \end{subfigure}
  \begin{subfigure}[b]{.23\linewidth}
    \includegraphics[page=2, width=\linewidth, trim={0 22.3cm 26.1cm 0}, clip]{plots_revision/page_status_small_pages.pdf}
    \vspace*{-0.3cm}
    \label{fig:async:w2:page_status}
  \end{subfigure}
   \begin{subfigure}[b]{.23\linewidth}
    \includegraphics[page=3, width=\linewidth, trim={0 22.3cm 26.1cm 0}, clip]{plots_revision/page_status_small_pages.pdf}
    \vspace*{-0.3cm}
    \label{fig:async:w3:page_status}
  \end{subfigure}
   \begin{subfigure}[b]{.23\linewidth}
    \includegraphics[width=\linewidth, trim={0 0.45cm 0 0}, clip]{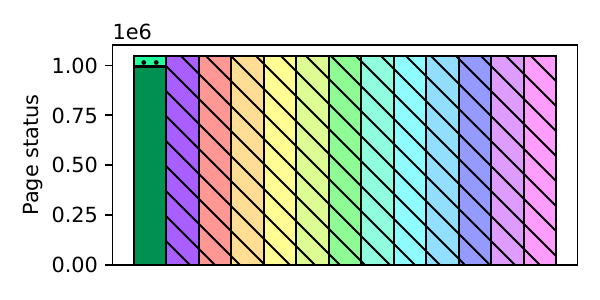}
    \vspace*{-0.3cm}
    \label{fig:async:w4:page_status}
  \end{subfigure}  
 
\vspace*{-0.2cm}
  \caption{Migration under concurrent writes for small pages and a reduction factor 2 for \m{}.}
    \label{fig:concurrent_writes:small_pages}
    \vspace*{-0.3cm}
\end{figure*}

\vspace*{-0.2cm}
\section{Migration Under Concurrent Writes}
\label{ssec:exp_asynchronous}
\vspace*{-0.1cm}

Next, we inspect the performance under concurrent writes, where for \m{}, potential additional costs from signal handling and retrying migrations comes into play. For each method, we asynchronously perform the migration while we fire a continuous burst of uniformly selected random writes on the memory, specified at a requested throughput in writes/s. As representatives for a low, high, and extreme pressure case, we test $10$K, $100$K, and $10$M (small pages) respectively $100$M (huge pages) writes/s. Additionally, we include a skewed workload where out of $100$K writes/s, 75\% concentrate only on $128$MB ($3.125\%$) of the memory. Firing the burst serves two purposes: First, we see how the migration of each method is negatively affected by the concurrent burst, i.e., by slowing down migration, increasing the number of retries, or even leaving pages un-migrated. Second, we see whether the migration also negatively impacts the write burst and at which point the requested throughput cannot be sustained anymore. Note that as soon as the migration finishes or a timeout occurs after 10s, the write burst also terminates and the experiment ends. We then also report the current location or error status of all pages.
For \m{}, we again show for completeness all previously tested area sizes, but focus on the most promising previously identified sizes of~$512$KB and $16$MB, where the stated area sizes are now \textit{initial} area sizes -- in combination with the used reduction factor of~$2$, the retry of an area now automatically splits it in half and retries the migration of the two halves. For auto NUMA balancing, since it does not signal its start or finish in any way, we check every $100$ms whether all pages have been migrated already and stop the measurement if yes. Otherwise, we continue until the timeout is reached.

\begin{figure*}[h!]
  \centering
  \raisebox{1.1cm}{\rotatebox{90}{\textbf{Migration time}}}
  \begin{subfigure}[b]{.23\linewidth}
  \centering \textbf{10K uniform writes/s}
    \includegraphics[width=\linewidth, trim={0 0.45cm 0 0}, clip]{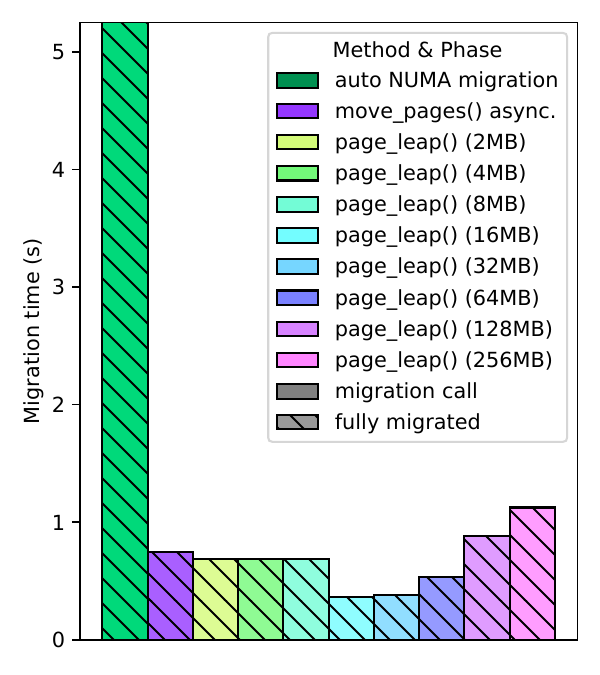}
    \vspace*{-0.3cm}
    \label{fig:sync:w1:migration}
  \end{subfigure}
  \begin{subfigure}[b]{.23\linewidth}
  \centering \textbf{100K uniform writes/s}
    \includegraphics[width=\linewidth, trim={0 0.45cm 0 0}, clip]{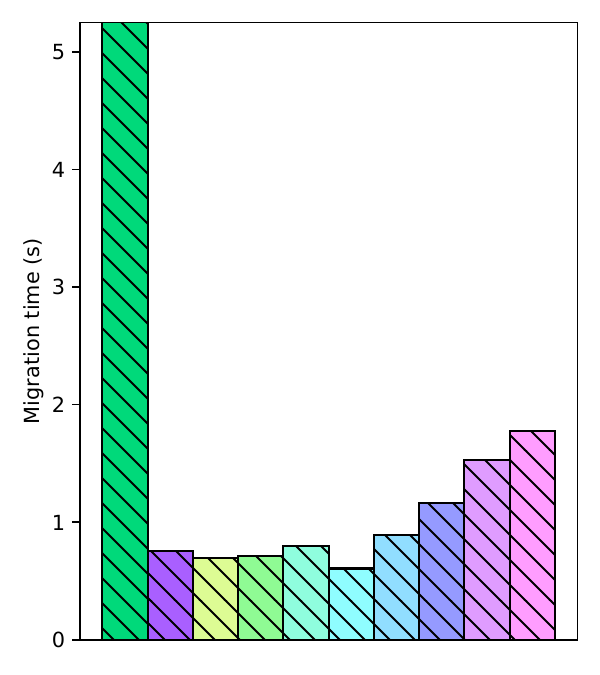}
    \vspace*{-0.3cm}
    \label{fig:sync:w2:migration}
  \end{subfigure}
   \begin{subfigure}[b]{.23\linewidth}
   \centering \textbf{100M uniform writes/s}
    \includegraphics[width=\linewidth, trim={0 0.45cm 0 0}, clip]{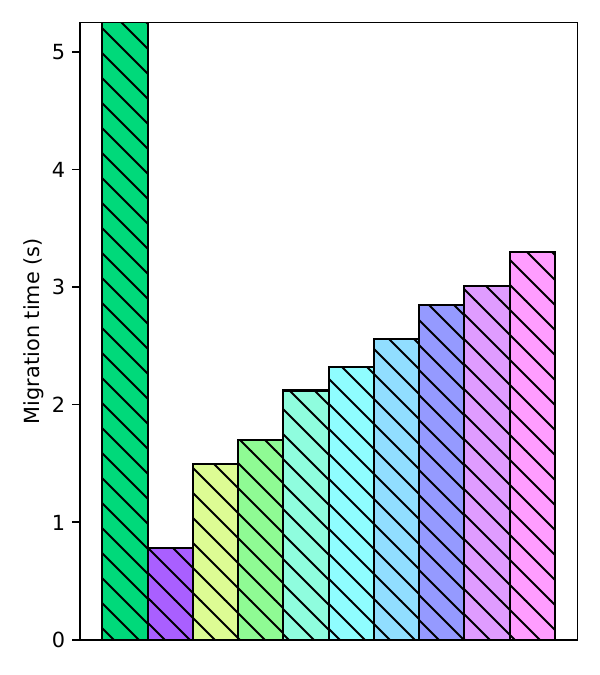}
    \vspace*{-0.3cm}
    \label{fig:sync:w3:migration}
  \end{subfigure}
   \begin{subfigure}[b]{.23\linewidth}
       \centering \textbf{100K writes/s, 75\% in 128MB}
    \includegraphics[width=\linewidth, trim={0 0.45cm 0 0}, clip]{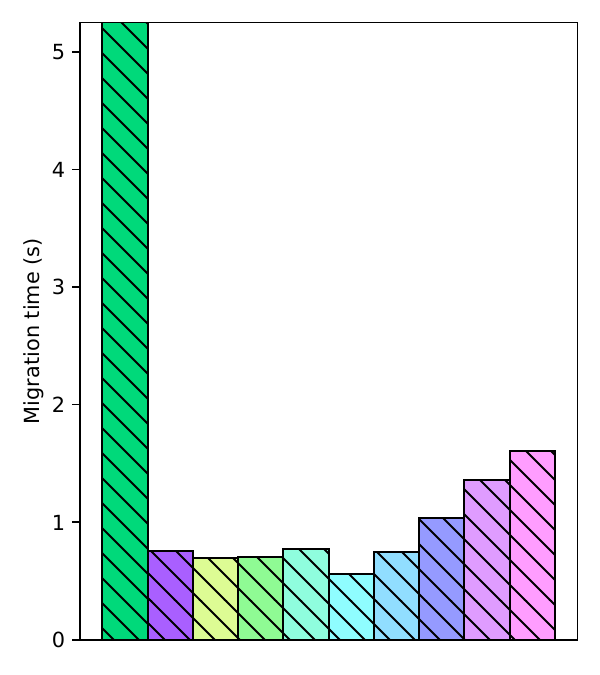}
    \vspace*{-0.3cm}
    \label{fig:sync:w4:migration}
  \end{subfigure}

\raisebox{0.1cm}{\rotatebox{90}{\textbf{Throughput}}}
  \begin{subfigure}[b]{.23\linewidth}
    \includegraphics[width=\linewidth, trim={0 0.45cm 0 0}, clip]{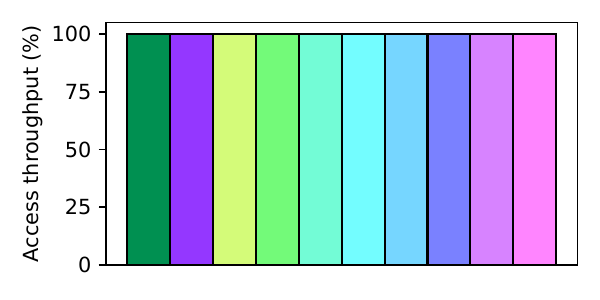}
    \vspace*{-0.3cm}
    \label{fig:sync:w1:access_throughput}
  \end{subfigure}
  \begin{subfigure}[b]{.23\linewidth}
    \includegraphics[width=\linewidth, trim={0 0.45cm 0 0}, clip]{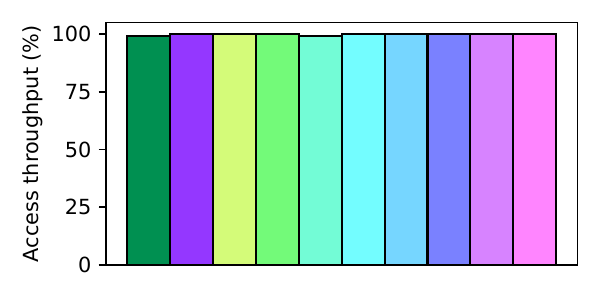}
    \vspace*{-0.3cm}
    \label{fig:sync:w2:access_throughput}
  \end{subfigure}
   \begin{subfigure}[b]{.23\linewidth}
    \includegraphics[width=\linewidth, trim={0 0.45cm 0 0}, clip]{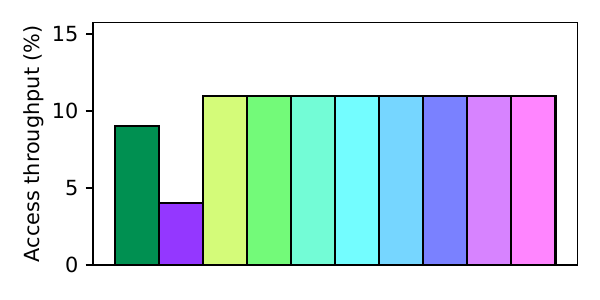}
    \vspace*{-0.3cm}
    \label{fig:sync:w3:access_throughput}
  \end{subfigure}
   \begin{subfigure}[b]{.23\linewidth}
    \includegraphics[width=\linewidth, trim={0 0.45cm 0 0}, clip]{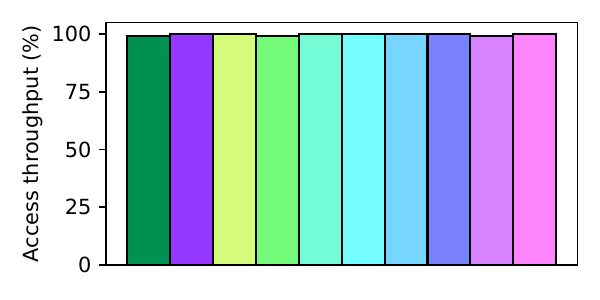}
    \vspace*{-0.3cm}
    \label{fig:sync:w4:access_throughput}
  \end{subfigure}  
  
\raisebox{0.1cm}{\rotatebox{90}{\textbf{Page status}}}
  \begin{subfigure}[b]{.23\linewidth}
    \includegraphics[width=\linewidth, trim={0 0.45cm 0 0}, clip]{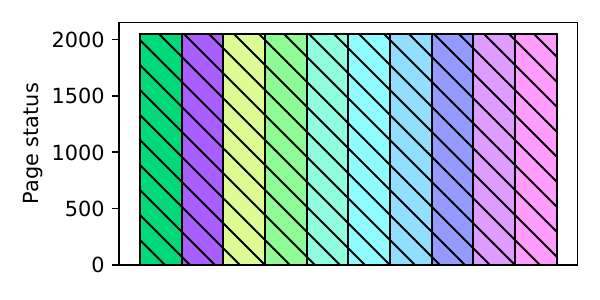}
    \vspace*{-0.3cm}
    \label{fig:sync:w1:page_status}
  \end{subfigure}
  \begin{subfigure}[b]{.23\linewidth}
    \includegraphics[width=\linewidth, trim={0 0.45cm 0 0}, clip]{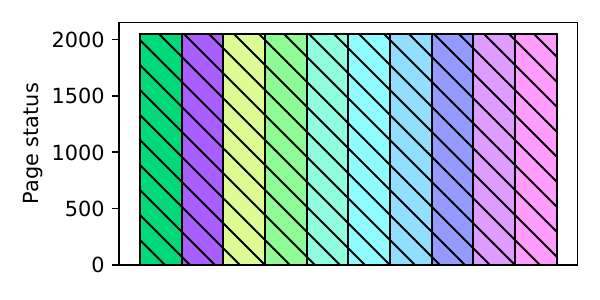}
    \vspace*{-0.3cm}
    \label{fig:sync:w2:page_status}
  \end{subfigure}
   \begin{subfigure}[b]{.23\linewidth}
    \includegraphics[width=\linewidth, trim={0 0.45cm 0 0}, clip]{plots_revision/page_migration_status_huge_pages_100000_writes_per_second_uniform.pdf}
    \vspace*{-0.3cm}
    \label{fig:sync:w3:page_status}
  \end{subfigure}
   \begin{subfigure}[b]{.23\linewidth}
    \includegraphics[width=\linewidth, trim={0 0.45cm 0 0}, clip]{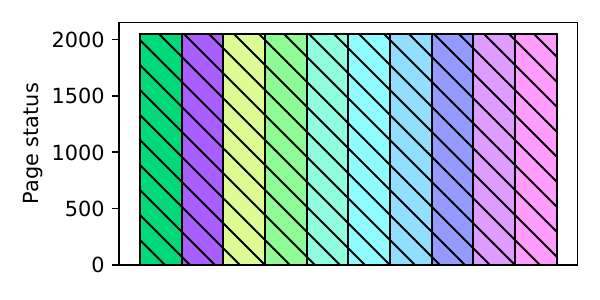}
    \vspace*{-0.3cm}
    \label{fig:sync:w4:page_status}
  \end{subfigure}  
 
\vspace*{-0.2cm}
  \caption{Migration under concurrent writes for huge pages and a reduction factor 2 for \m{}.}
    \label{fig:concurrent_writes:huge_pages}
    \vspace*{-0.3cm}
\end{figure*}

We first look at the results for small pages in Figure~\ref{fig:concurrent_writes:small_pages}, where we report the migration time, the actually achieved throughput as percentage of the requested throughput, and the page status.
When comparing the migration time first, we can see that for all workloads, \m{} manages to significantly outperform both \smalltt{move\_pages()} and auto NUMA balancing for a large number of initial area sizes. The previously determined $16$MB clearly performs best for the $10$K and $100$K writes/s cases, whereas for the extreme case of $10$M requested writes/s, an initial area size of $512$KB performs best. For the skewed case, both area sizes perform equally well. Hence, for small pages, we recommend to generally run \m{} with an initial area size of $16$MB to hit the sweet spot --- under normal pressure, we then exploit the effectiveness of larger copies and fewer remappings, while under extreme pressure, the method is still able to quickly enough adapt to smaller area sizes to still outperform the alternatives.  
When inspecting the achieved throughput of writes, we can see that all methods handle the low and high pressure cases well and stay close to 100\%. Only under extreme pressure, the throughput cannot be sustained anymore.
Regarding the page status, \m{} (for reasonable initial area sizes) and \smalltt{move\_pages()} always manage to migrate all pages before the timeout occurs. Auto NUMA balancing performs poorly for all workloads, as it migrates only a small portion of the pages.
For huge pages in Figure~\ref{fig:concurrent_writes:huge_pages}, we can see that for the low pressure, high pressure, and skewed case, \m{} for an initial area size of $16$MB still outperforms \smalltt{move\_pages()} by up to $2$x, but the advantage is overall smaller. Due to the larger underlying page size, all methods now operate closer to the theoretical copying optimum. We have to admit that \smalltt{move\_pages()} shows a more robust performance for the extreme case --- however it also achieves a lower throughput. Also, although that auto NUMA balancing times out, at the end, all pages have been migrated. This shows that for huge pages, where a smaller number of pages is in use, auto NUMA balancing utilizes the ``pause'' after the write burst has ended and actually migrates the pages. Still, it again shows its unpredictability in terms of when and if at all the migration actually happens.

Next, in Table~\ref{tab:overhead}, we quantify the actual overhead of \m{} under concurrent writes for the case of $100$K writes/s. To do so, we first measure for different area sizes the total amount for memory \m{} actually has to copy, including the memory additionally copied during retries. Then, we measure how long it takes to copy the same amounts of memory via \smalltt{memcpy()} from one region to the other. Here, we make sure to also copy areas individually. We then report the difference in time, which essentially resembles the overhead of \m{} for everything that happens on top of the raw copying. Table~\ref{tab:overhead} shows the memory overhead and time overhead for both page sizes, where we include the (unpractical) extremes of $4$KB and $256$MB as well as the recommended initial area sizes of $512$KB and $16$MB. For the $4$KB case, we can clearly see that the system call overhead is the bottleneck, as no retries are happening at all. For $256$MB, around $420\%$ are additionally copied, leading to a $27.5\%$ respectively $18.3\%$ time overhead. The recommended sizes however manage to keep both the memory and time overhead at most around $50\%$ over \smalltt{memcpy()}, which shows the efficiency of \m{} for reasonably chosen initial area sizes.


\begin{table}[h!]
\vspace*{-0.2cm}
    \footnotesize
    \centering
    \begin{tabular}{c|c|c|c|c}
    \toprule
        & \multicolumn{2}{c|}{\textbf{Small pages}}  & \multicolumn{2}{c}{\textbf{Huge pages}} \\
        \midrule
        \textbf{Area} & \textbf{Memory} & \textbf{Time} & \textbf{Memory} & \textbf{Time} \\
    \midrule
       4KB & 0MB (0.0\%)& 5635ms (659\%) & - & - \\\hline
       \textbf{512KB} & \textbf{5.5MB (0.13\%)}& \textbf{210ms (31.3\%)} & - & - \\\hline
       2MB & 42MB (1.02\%) & 155ms (22.9\%) & 66MB (1.61\%) & 34ms (5.1\%)\\\hline
       \textbf{16MB} & \textbf{2.1GB (52.5\%)} & \textbf{245ms (47.5\%)} & \textbf{1.8GB (45\%)} & \textbf{135ms (28.4\%)}\\\hline
       256MB & 17.1GB (427.5\%) & 437ms (27.5\%) & 16.80GB (420\%) & 274ms (18.3\%)\\
    \bottomrule
    \end{tabular}
    \caption{Time overhead and memory overhead over \smalltt{memcpy()}.}
    \label{tab:overhead}
    \vspace*{-0.5cm}
\end{table}

\begin{figure}[h!]
    \centering
    \includegraphics[page=2, width=\columnwidth, trim={0 13cm 0.2cm 0}, clip]{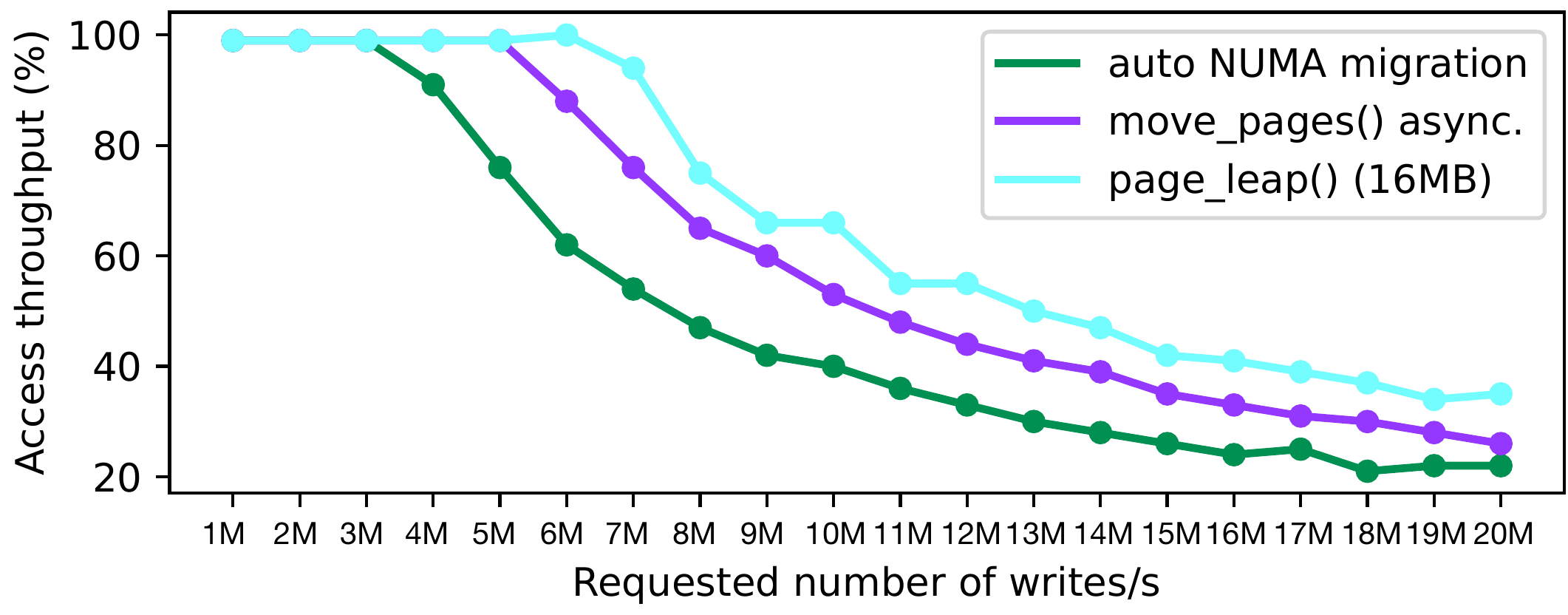}
    \caption{Achieved write/s throughput for a run of 10s.}
    \label{fig:no_early_abort}
    \vspace*{-0.5cm}
\end{figure}

So far, we have analyzed the access throughput purely while the migration is happening by stopping the fired burst as soon as the migration has ended (or the timeout occurs). To now also measure the positive impact of a fast migration, namely being able to perform local accesses earlier, we adjust the experiment to fire the write burst for the whole 10s, independent of when the migration finishes. Figure~\ref{fig:no_early_abort} shows the results for small pages while varying the requested throughput. We can see that when increasing the requested throughput, \m{} is able to sustain it significantly longer than both baselines and constantly performs best: For instance, at $6$M requested writes/s, \m{} can still fully sustain the throughput, while auto NUMA balancing achieves only 65\% of it. This shows that migrating as fast as possible significantly pays off as less remote accesses happen over the total duration.

\vspace*{-0.2cm}
\section{Handling Database Workloads}
\label{sec:tpch}
\vspace*{-0.1cm}

Finally, let us evaluate how \m{} handles actual database workloads. We evaluate a scenario as it could occur in a morsel-driven database engine~\cite{lit:morsel} that semantically partitions its database across two NUMA regions, where a thread of one region is currently idling, while all morsels that still need to be processed are located on the other region. In this situation, the thread can migrate the morsels over to region~1 to speed up the query processing. 

\begin{figure}[h!]
  \centering
  \begin{subfigure}[b]{.49\linewidth}
    \includegraphics[width=\linewidth, trim={0 0 0 0}, clip]{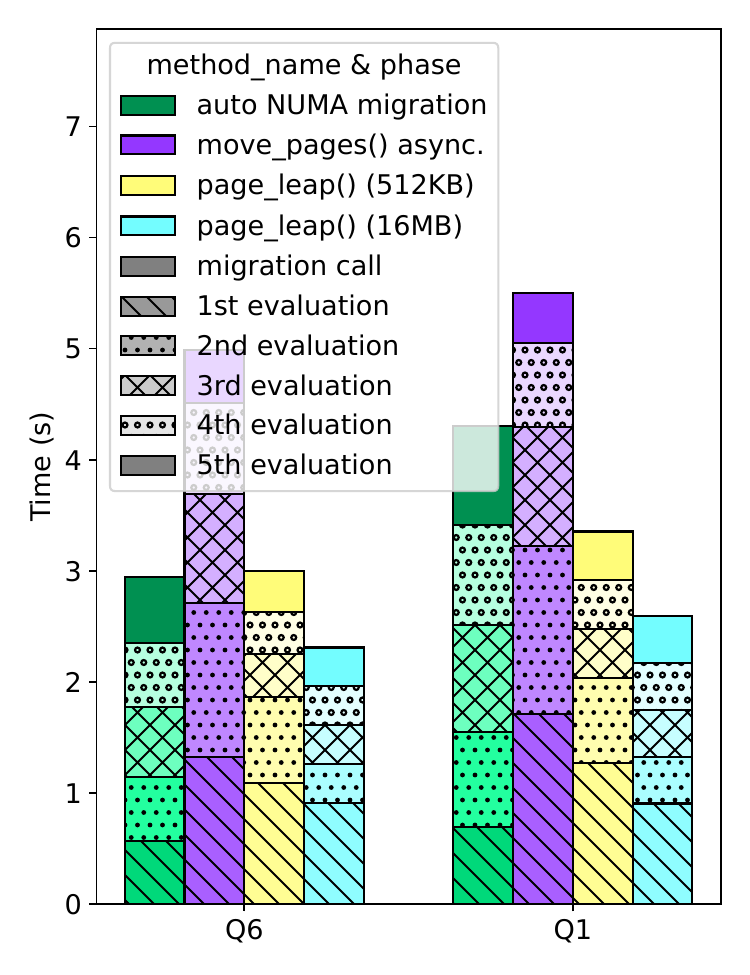}
    \caption{Without concurrent writes.}
    \label{fig:tpch:no_writes}
  \end{subfigure}
  \begin{subfigure}[b]{.49\linewidth}
    \includegraphics[width=\linewidth, trim={0 0 0 0}, clip]{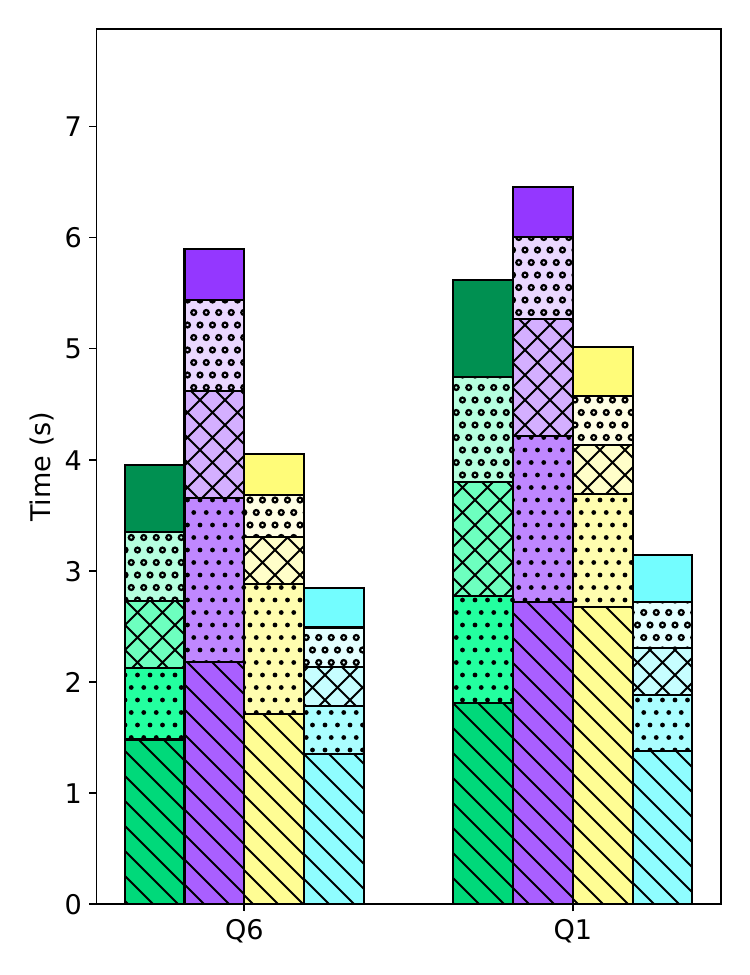}
    \caption{With concurrent writes.}
    \label{fig:tpch:writes}
  \end{subfigure}
\vspace*{-0.5cm}
  \caption{TPC-H workloads.}
    \label{fig:tpch}
    \vspace*{-0.3cm}
\end{figure}

To analyze this scenario, we allocate a set of morsels of a total size of~$1$GB of the TPC-H \smalltt{lineitem} table on NUMA region~0. We then migrate the morsels over via ~\smalltt{move\_pages()} respectively ~\m{} into pooled memory before query answering. Additionally, we include auto NUMA balancing. As queries, we evaluate hand-written variants of TPC-H Q1 and Q6, as they operate on \smalltt{lineitem} only and study their performance with and without concurrent writes fired without delay, which are carried out by a separate thread that performs $10M$~writes into the \smalltt{L\_ORDERKEY} field of uniformly selected rows. As \smalltt{L\_ORDERKEY} is neither used by Q1 nor Q6, this does not impact the query result, but still penalizes the migration procedures.   
Figure~\ref{fig:tpch} shows the time for the asynchronous migration call (for \m{} and \smalltt{move\_pages()}) followed by a sequence of five executions of the respective query. We can observe that \m{} again clearly outperforms \smalltt{move\_pages()}, where \m{} pays off over \smalltt{move\_pages()} already from the second query on for both recommended area sizes. This shows that \m{} can also operate effectively in the context of database workloads. 

\vspace*{-0.3cm}
\section{Related Work}
\label{sec:related_work}
\vspace*{-0.1cm}

Before concluding, we discuss the related work in the field. 

First, in terms of general-purpose techniques for NUMA migration, other built-in methods are worth mentioning: One representative is \smalltt{numactl}, a command-line tool by which page migration can be triggered as well. However, as it cannot be comfortably used from \textit{within the process} like the remaining methods, we did not consider it as a baseline here. Also, there exists the system call \smalltt{migrate\_pages()} from \smalltt{libnuma}. However, in contrast to \smalltt{move\_pages()}, it is not able to migrate only a subset of pages of the process, but always migrates all pages, which is typically not flexible enough. Hence, it does not provide a fine-grained way of page migration.  
Related to \smalltt{move\_pages()} is also the recent work of~\cite{lit:move_pages_2}, which presents a faster version of the system called \smalltt{move\_pages2()}. However, as this requires modifying the kernel while \m{} works out-of-the-box on a vanilla kernel, we did not consider it as a baseline here. 
In the database context, NUMA-aware algorithms and data structures have been a topic of interest for many years. These works operate around efficient data placement~\cite{lit:dbms_numa_1} and scheduling~\cite{lit:dbms_numa_4} on NUMA systems, NUMA-aware joins~\cite{lit:dbms_numa_2}, and NUMA-aware execution engines from a holistic perspective~\cite{lit:dbms_numa_3}.
In terms of memory rewiring, numerous other works also exist that exploit it. These range from data structures, such as linked lists~\cite{lit:rewiring}, packed memory arrays~\cite{lit:packed_memory_arrays}, resizable hash tables schemes~\cite{lit:shortcut}, and radix trees~\cite{lit:start} to algorithms such as out-of-place partitioning~\cite{lit:rewiring} and snapshotting~\cite{lit:anyolap, lit:no_time_to_halt}.  

\vspace*{-0.3cm}
\section{Conclusion}
\label{sec:conclusion}
\vspace*{-0.1cm}

In this paper, we analyzed the weaknesses of the built-in implicit and explicit page migration mechanisms, auto NUMA balancing, and \smalltt{move\_pages()}, and proposed a practical alternative that works in user-space on a vanilla Linux kernel. We experimentally demonstrated that our alternative significantly reduces migration overhead while ensuring correctness under concurrent writes and providing a guarantee to migrate eventually. Further, our method supports the migration into pooled memory, which offers a significant advantage over the baseline methods.

\newpage
\section*{Artifacts}

All code of the project, including build instructions and all artifacts, is available under \url{https://gitlab.rlp.net/fschuhkn/efficient_numa_page_migration}.

\bibliographystyle{ACM-Reference-Format}
\bibliography{bibliography}

\end{document}